# Spin-Torque Ferromagnetic Resonance Measurements of Damping in Nanomagnets

G. D. Fuchs, J. C. Sankey, V. S. Pribiag, L. Qian, P. M. Braganca, A. G. F. Garcia, E. M. Ryan, Zhi-Pan Li, O. Ozatay, D. C. Ralph, and R. A. Buhrman.

Abstract

We measure the magnetic damping parameter $\alpha$ in thin film CoFeB and permalloy (Py) nanomagnets at room temperature using ferromagnetic resonance driven by microwave frequency spin-transfer torque. We obtain $\alpha_{CoFeB} = 0.014 \pm 0.003$ and $\alpha_{Py} = 0.010 \pm 0.002$, values comparable to measurements for extended thin films, but significantly less than the effective damping determined previously for similar nanomagnets by fits to time-domain studies of large-angle magnetic excitations and magnetic reversal. The greater damping found for the large amplitude nanomagnet dynamics is attributed to the nonlinear excitation of non-uniform magnetic modes.

Dissipation in nanoscale magnetic systems is of wide-spread fundamental interest due to its role in the magnetization dynamics of spatially confined magnetic elements[1]. In addition, understanding and controlling magnetic damping is important for minimizing the switching current in proposed future generations of magnetic memory switched by spin-torque[2] (ST) and for counter-acting ST-excited magnetic noise, which may limit the areal density in future generations of hard-drives that use giant magnetoresistance (GMR) read-heads[3]. Magnetic damping is usually characterized by the phenomenological damping parameter $\alpha$, which can have both intrinsic and extrinsic contributions. The latter includes surface effects, which may be particularly important in patterned magnetic nanostructures. Furthermore, in multilayer structures, the pumping of spins from a precessing magnetic moment can also cause additional damping[4], an effect which may depend on the amplitude of the magnetic excitation[5].

Although conventional damping measurement techniques cannot be readily applied to individual nanoscale structures, experiments that employ spin-transfer torque to control magnetic dynamics can measure damping in nanomagnets via several approaches. In previous work, time-domain measurements of coherent relaxation oscillations of a $Ni_{81}Fe_{19}$ (Py) layer in a nanopillar spin valve device that was excited by spin torque from a short current pulse gave $\alpha = 0.025$ at 40 K[6]. Macrospin modeling of short-pulse ST-switching experiments has also yielded self-consistent and unique results for the damping and spin-transfer efficiency over a broad range of experimental parameters.[7,8] These fits gave values of $\alpha$ for Py nanomagnets at room temperature (RT) of 0.030 – 0.035, which are much larger than those obtained from conventional damping measurements on extended thin films[9,10,11] (0.006-0.012). An analysis of pulse-switching measurements made at low temperature (LT) yielded even higher values[8], $\alpha \geq 0.05$. While a LT increase in damping can be attributed to the presence of an adventitious oxide around the perimeter of the nanomagnet that is cooled below the oxide's antiferromagnetic (AF) blocking temperature, the large RT values of $\alpha$ cannot readily be ascribed to either the native oxide or to spin pumping. Since these differing prior results are limited to relatively large-angle magnetization dynamics and to magnetic reversal, they motivate the development of small-angle techniques for direct measurements of nanomagnet dynamics in the field and temperature regime that is of primary interest for many magnetoelectronic applications.

Here we demonstrate that this can be accomplished by using spin-transfer-driven ferromagnetic resonance (ST-FMR).[12,13,14,15] We have determined the magnetic damping for small-angle precession in spin valve nanopillars composed of Py/Cu/Py as well as CoFeB/Cu/CoFeB and find $\alpha$ = 0.010±0.002 and 0.014±0.003, respectively. The results demonstrate that the nanofabrication processes used to form magnetic nanopillars do not necessarily increase ferromagnetic damping. They also suggest that the larger values of effective damping determined previously by modeling RT ST switching are associated with the large-amplitude magnetic oscillations intrinsic to that process, and possibly to the loss of energy to spin wave modes that do not contribute to switching.

We studied two types of magnetic multilayers (in nm): Ta 4/Cu 22/Ta 5/Cu 22/Ta 20/CoFeB 20/Cu 6/CoFeB 3.5/Cu 5/Pt 30 and Py 4/Cu 120/Py 20/Cu 12/Py 5.5/Cu 20/Pt 30. We will refer to the thicker magnetic layer in each device as the "fixed layer" and the thinner layer as the "free layer". All films were deposited on thermally oxidized silicon wavers via dc magnetron sputtering at room temperature in a vacuum system with a base pressure of $3\times10^{-8}$ torr. CoFeB was sputtered from an alloy target with atomic ratios of 60/20/20 and had a RT magnetization ($4\pi M$) of 14.8 kOe, while the Py had $4\pi M$=7.0 kOe. These multilayers were then patterned using electron-beam lithography and ion milling to form nanopillar spin-valve structures, with approximately elliptical cross sections having the nominal dimensions 50×110 nm for the CoFeB device and 55×130 nm for the Py device. Our maximum processing temperature was 170º C, so we expect that the CoFeB remained in the amorphous state after fabrication into nanopillars[16].

Figure 1 (a) shows a schematic of the ST-FMR measurement setup, which is similar to that used in Ref. 13. A microwave current $I_{RF}$, pulsed with a ~1.3 kHz repetition rate, is applied perpendicular to the layers of the nanopillar to generate a microwave-frequency spin-transfer torque, which can excite precession in the magnetization of the either the fixed and free magnetic layers (or both) when the drive is resonant with a magnetic normal mode. A direct current $I_{dc}$ can be applied simultaneously via a bias-tee. In order for the spin-transfer torque to be non-zero, the magnetizations of the fixed and free magnetic layers must be misaligned from either the strictly parallel or anti-parallel configuration. In this experiment, we induce misalignment by applying an in-plane external field ($H_{appl}$) at a large angle with respect to the nanopillar easy axis[12] (Fig. 1 (b)). Through the GMR effect, magnetic precession in the multilayer generates an ac resistance that mixes with $I_{RF}$ to produce a rectified voltage, $V_{mix}$, which is detected with a lock-in amplifier.

The misalignment angle is determined by the in-plane uniaxial anisotropies, $H_k$, of the free and fixed layers for each sample, which can be estimated based on GMR measurements. For the CoFeB sample we estimate $H_{k,free}$= 850 Oe and $H_{k,fix}$= 700 Oe, and for the Py sample $H_{k,free}$=430 Oe and $H_{k,fix}$= 440 Oe[17].

To measure the resonance linewidth, we apply constant microwave power to the sample and measure $V_{mix}$ vs. $f$ at different values of $I_{dc}$. The effect of spin transfer from $I_{dc}$ is to decrease the effective damping as $I_{dc}$ is stepped to negative values, so that the resonant response to $I_{RF}$ grows and the signal amplitude becomes larger as $I_{dc}$ decreases toward the critical current. Figure 2(a) shows a representative ST-FMR peak for the CoFeB sample, measured with $I_{RF}$=0.18 mA and $I_{dc}$ = -2.0 mA, and with $H_{appl}$ = 200 Oe at an 80º angle with respect to the x-axis. We estimate the GMR angle $\theta$ to be ~163º and that the free layer magnetization rotates to an angle $\phi$ of ~ 177º under the influence of $H_{tot}$=493 Oe along an angle $\psi$=172º relative to the x-axis (Fig. 1 (b)). $H_{tot}$ is the vector sum of $H_{appl}$ and the dipolar field of the fixed layer (Fig. 1(b).). We have confirmed that the linewidth and lineshape remain the same for smaller microwave currents, indicating that we are operating in the linear-response regime. Based on the signal amplitude, the precessional angle for the $I_{dc}$=2.0 mA data is ~6º, and for all other values of DC current that we report for the CoFeB sample the precession angle is less than or equal to this value.

To determine the magnetic damping, we fit the ST-FMR lineshapes to a combined symmetric and anti-symmetric Lorenztian of the form:

$$\frac{A}{1+(f-f_o)^2/\Delta_o^2} - \frac{B(f-f_o)/\Delta_o}{1+(f-f_o)^2/\Delta_o^2}. \tag{1}$$

An anti-symmetric component of the ST-FMR line-shape can arise when the spin-torque vector is not in a principal plane of the anisotropy tensor,[14,15] or from contributions of an "effective field" component of spin torque perpendicular to the magnetizations of both magnetic layers[12]. For our metal spin-valve samples we find a small |B/A| ratio between 0 and 0.07, which varies slightly among different samples, presumably due to anisotropy and shape variations. The damping $\alpha$ is related to the FMR half-width $\Delta_o$ by:[15]

$$\alpha = \frac{\Delta_o}{\Delta_k}, \text{ where } 2\pi\Delta_k = \gamma 4\pi M(N_y' + N_z')/2 \text{ and} \tag{2}$$

$$N_y' = (N_y - N_x)(\cos^2\phi - \sin^2\phi) + \cos(\psi-\phi)H_{tot}/(4\pi M) \tag{3}$$

$$N_z' = (N_z - N_x\cos^2\phi - N_y\sin^2\phi) + \cos(\psi-\phi)H_{tot}/(4\pi M). \tag{4}$$

where $N_x$, $N_y$, $N_z$ are the demagnetization factors. Eqs. (3) and (4) describe the effective in-plane and out-of-plane anisotropy factors, respectively. For our samples, the determination of the damping from the FMR half-width is dominated by the out-of-plane anisotropy since $N_x \approx N_y \approx 0$ and $N_z \approx 1$ in the thin-film limit. In determining the damping quantitatively, we use the demagnetization factors calculated from the sample geometry assuming that the free layers are elliptical cylinders[18]. We find for the CoFeB free layer: $N_x$=0.034, $N_y$=0.091, $N_z$=0.876, and for the Py free layer, $N_x$=0.044, $N_y$=0.105, $N_z$=0.851. These numbers are consistent with the observed FMR frequencies as well as with 4 K coercive field measurements in multiple samples with the same geometry.

Figure 2 (c) shows $\alpha$ as a function of $I_{dc}$, determined from this fitting procedure. We find that $\alpha$ depends linearly on $I_{dc}$ as expected for the lowest-frequency free layer mode, because spin transfer from $I_{dc}$ should modify the effective damping[19]. The regression line gives $\alpha$ = 0.014±0.003 at $I_{dc}$ = 0. We estimate the error by propagating uncertainty in the determination of the anisotropies, the magnetization angles as well as from fits to the data. In addition, when we perform damping measurements with an initial magnetization state misaligned from the parallel (rather than the antiparallel) configuration, we find the same value of α within the experimental accuracy.

We measured the Py sample at $H_{appl}$ = 200 Oe with at a 70º angle from the easy axis, which induces a ~154º GMR angle with $\phi$ ~172º, $\psi$=158º and $H_{tot}$ = 245 Oe. Fig. 2(b) shows the resonant response with $I_{RF}$ = 0.035 mA and $I_{dc}$ = -0.5 mA, and Fig. 2(d) shows α as a function of $I_{dc}$. As before, we observe a linear trend in α as we step $I_{dc}$. At $I_{dc}$ = 0, we find $\alpha$ = 0.010±0.002. The maximum precession angle in these data is ~6.5º at $I_{dc}$ = -0.7 mA.

The values of damping that we obtain from the ST-FMR measurements are quite consistent with the results obtained for CoFeB and Py extended thin-film multilayers using either field FMR or time-resolved techniques, 0.006-0.013 for CoFeB[20] and 0.006-0.012 for Py[9,10,11]. That the ST-FMR measured values of $\alpha$ are on the high end of these ranges can be attributed to the modest enhancement of damping that is expected to be present due to spin-pumping[4,10] given the proximity of the Pt capping layer separated from the free layer by 20 nm (5 nm) of Cu in the Py (CoFeB) sample.

Since the values of $\alpha$ determined by the ST-FMR are in accord with standard thin film FMR measurements, they are much less than the effective damping parameters 0.030-0.035 obtained from fits to short-pulse spin-transfer-driven magnetic switching experiments[7]. However, the fact that fits to pulse-switching data has been successful in describing the observed switching behavior over a broad range of pulse durations and amplitudes with a single set of torque and damping parameters is strong evidence of its validity for switching measurements, despite disaccord with our ST-FMR measurements. This disagreement can be understood in light of the difference between the two measurement regimes. ST-FMR damping measurements probe the energy dissipation of a single, small-amplitude excitation mode, while the switching experiments probe samples much farther from equilibrium, where nonlinear magnetic effects become important. In short-pulse ST switching experiments, if the spin-wave mode (or modes) that leads to reversal becomes non-linear at large amplitude, it can excite other magnetic modes. If these other modes do not directly support the reversal process, this energy transfer will have a similar effect as an increase in intrinsic damping. As additional evidence for this explanation, we note that similar magnetic nanopillar samples at low $T$ exhibit steady-state DC-driven magnetic precession over a range of current bias that is considerably broader than predicted by macrospin simulations[21]. The stability of these relatively large oscillations is most easily explained by an effective magnetic damping that increases with precession amplitude.

In summary, we have presented ST-FMR measurements of magnetic damping for small-angle magnetic precession in spin valve nanopillars. For devices with CoFeB magnetic layers, we find $\alpha=0.014\pm0.003$ and for Py, we find $\alpha=0.010\pm0.002$. These values are consistent with measurements made on continuous films, which demonstrates that processes for fabricating nanoscale structures do not necessarily lead to increased damping at room temperature. The considerably larger values of effective damping that have been obtained by fitting short-pulse switching data to macrospin models are attributed to the nonlinear transfer of spin angular momentum to spin wave modes that do not effectively contribute to magnetic reversal. The suppression of the excitation of such extraneous modes by the fabrication of smaller, magnetically more uniform and geometrically more ideal structures could substantially reduce the current amplitudes required for high speed ST switching.

This research was supported in part by the NSF/NSEC program through the Cornell Center for Nanoscale Systems and by the Office of Naval Research. Additional support was provided by NSF through use of the Cornell Nanoscale Science and Technology Facility/NNIN and the facilities of the Cornell Center for Materials Research.

Figure Captions:

Fig. 1. (a) Schematic diagram of the apparatus. (b) Left: A top-view SEM image of the an e-beam defined etch mask similar to the one used to form the CoFeB sample. The arrows on the SEM image represent the approximate orientation of $M_{free}$ and $M_{fixed}$ in our measurement, and the angle $\theta$ between them is the GMR angle. $H_{appl}$ marks the direction of the applied external field. Right: Coordinate system used in our analysis. $H_{tot}$ is the total field vector, composed of the vector sum of $H_{appl}$ and the dipolar field of the fixed layer, oriented along an angle $\psi$ in the film plane. $\phi$ is the angle of $M_{free}$.

Fig. 2. (a) ST-FMR data for the CoFeB sample taken at $H_{appl}$=200 Oe, $I_{RF}$=0.18 mA, and $I_{dc}$=-2.0 mA. The solid line is a Lorentzian fit. (b) Plot of the effective damping $\alpha$ vs. $I_{dc}$ for the CoFeB sample. The dashed line is a linear fit. (c) ST-FMR data for the Py sample taken at $H_{appl}$=200 Oe, $I_{RF}$=-0.035 mA, and $I_{dc}$=-0.5 mA. (d) Effective damping for the Py sample.

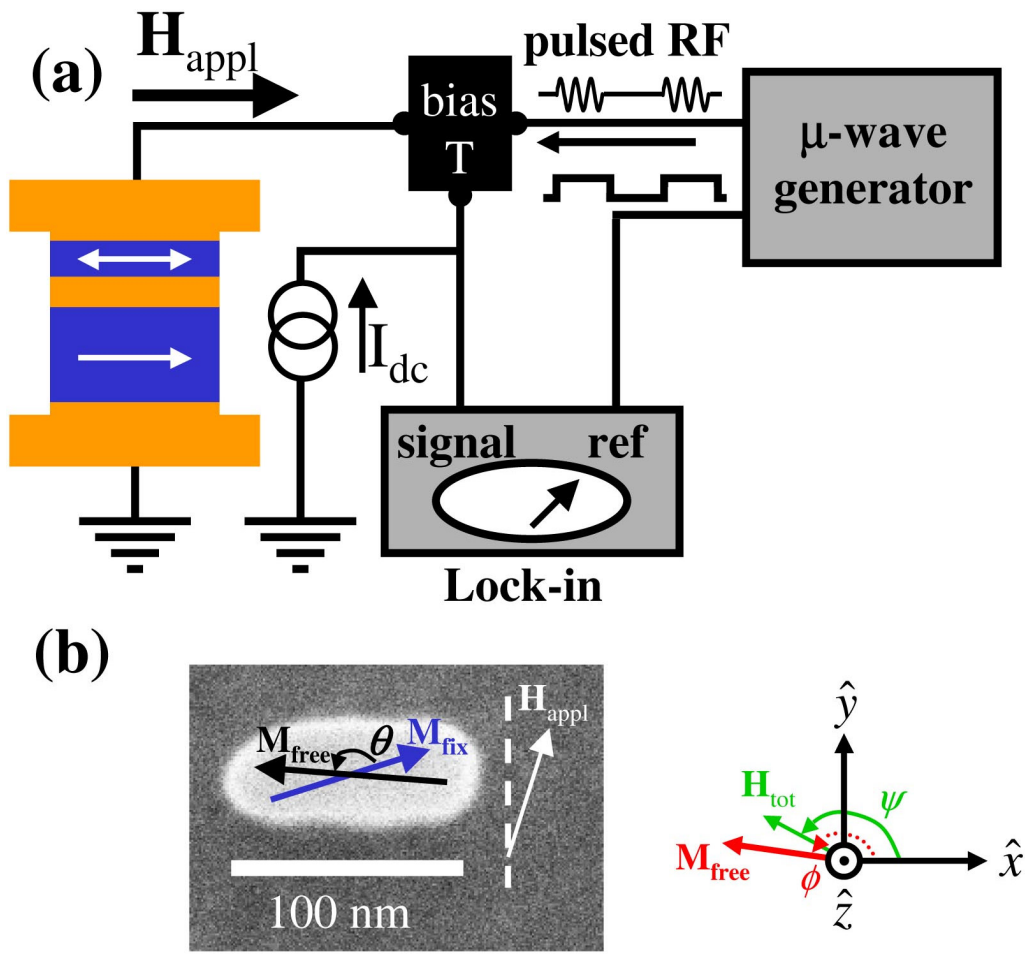

Fuchs *et. al,* Figure 1

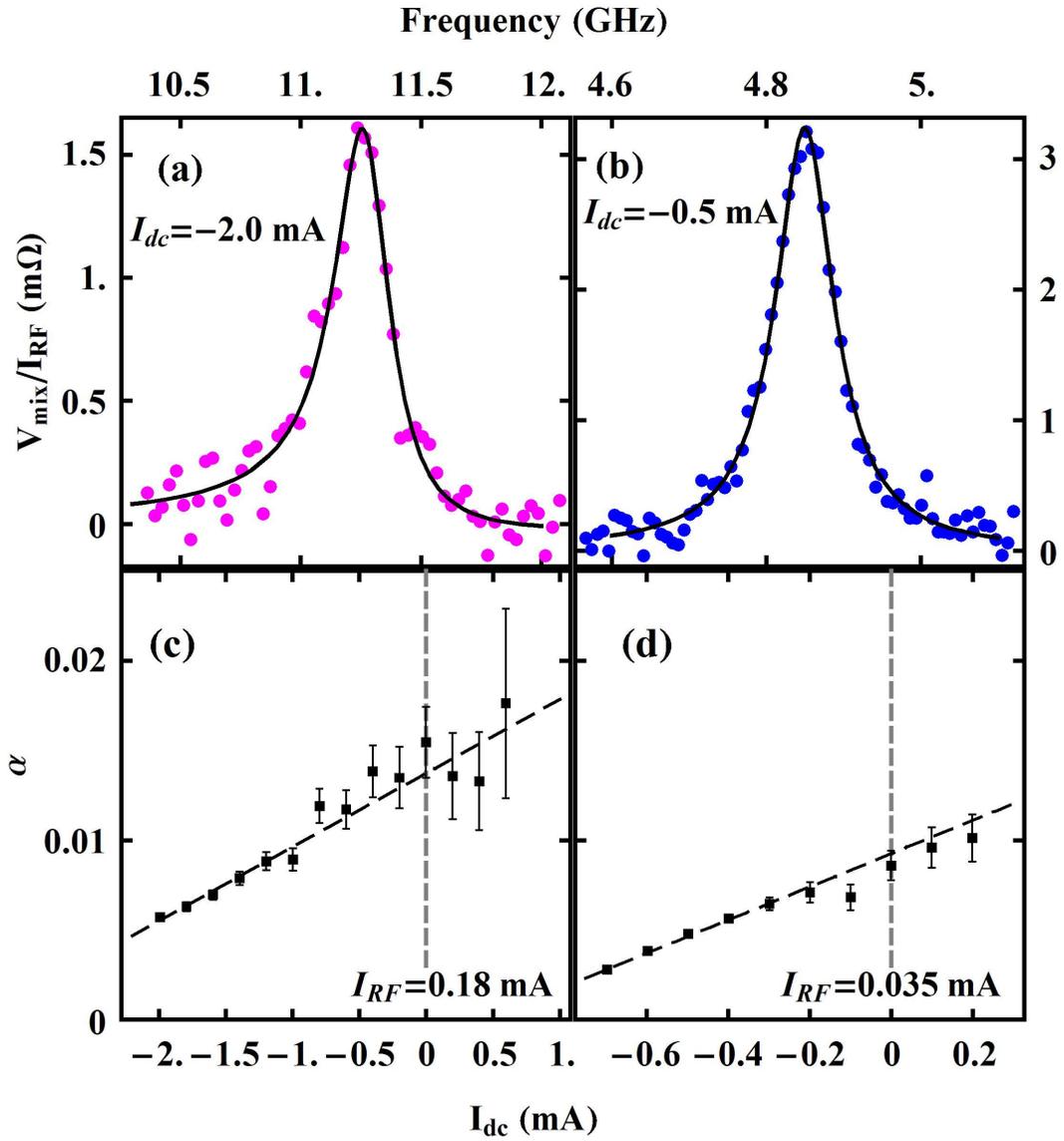

Fuchs *et. al*, Figure 2